\documentclass[12pt,preprint]{aastex}

\usepackage{epstopdf}
\DeclareGraphicsExtensions{.pdf,.eps,.png,.jpg,.mps}

\newcommand{\kms}{$\mathrm {km\,s}^{-1}$}
\newcommand{\mdot}{$\mathrm {M_{\odot}\,yr}^{-1}$}

\shorttitle{Flows in the interstellar medium}
\shortauthors{C. J. Wareing}

\begin{document}

\title{Reconciling the emission mechanism discrepancy in Mira's tail,
and its evolution in an interface with shear}

\author{C. J. Wareing\altaffilmark{1}}

\altaffiltext{1}{School of Mathematics, University of Leeds, 
Leeds, LS2 9JT, UK; C.J.Wareing@leeds.ac.uk}

\begin{abstract}

GALEX observations of the Mira AB binary system revealed a surrounding 
structure that has been successfully hydrodynamically interpreted as a 
bow shock and tail of ram-pressure-stripped material. Even the narrow 
tail, initially difficult to model, has been understood as the effect of the passage of
Mira from a warm neutral medium into a hot, low-density medium, 
postulated to be the Local Bubble. However, no model to date has 
explained the observed kink and associated general curvature of the tail. 
We test the hypothesis that before entering the Local Bubble, 
Mira was travelling through a shear flow with approximately 1/3 Mira's 
own velocity at an angle of ~30$^\circ$ to Mira's proper motion. 
The hypothesis reproduces the kinked nature of Mira's tail and predicts 
recompression and reheating of the tail material to the same or greater levels 
of density and temperature predicted in the shock. This provides a heat
source for the FUV emission, allowing for an extended lifetime of the FUV 
emission in line with other estimates of the age of the tail. 
The uniqueness of Mira's situation implies that the chances of observing other 
FUV tails behind AGB stars is highly unlikely.
 
\end{abstract}


\section{Introduction}

\begin{figure}
\begin{center}
\epsscale{0.275}
\plotone{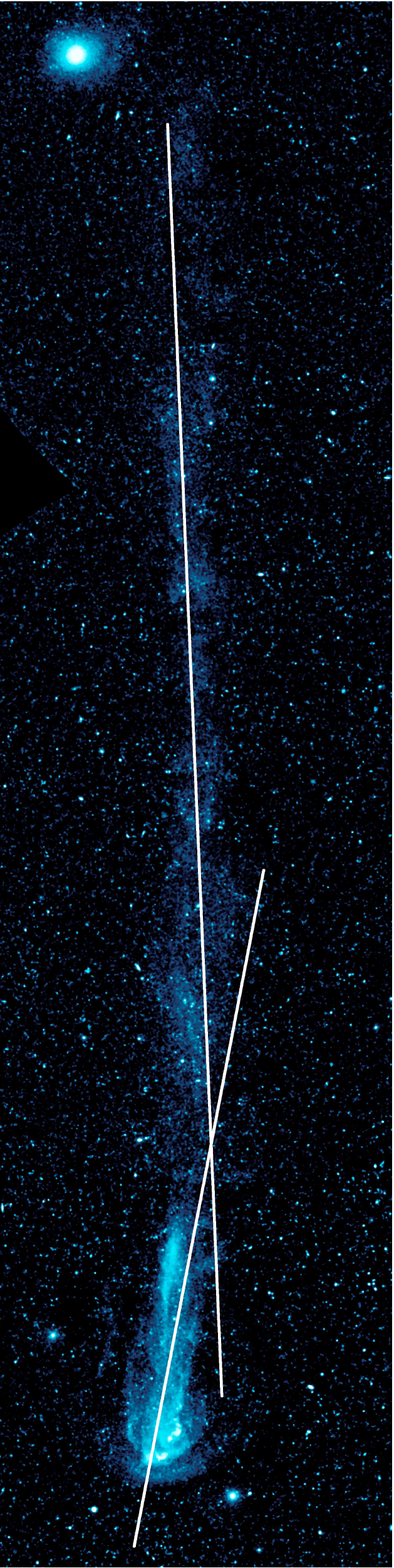}
\caption{A mosaic'd UV images of the Mira AB bow shock and tail system. 
This figure is an adaptation of figure 1(a) from \cite{martin07}. The two lines 
indicate the change of alignment in the tail. The bright star close to the 
termination point of the tail is HR 691.} 
\label{observation}
\end{center}
\end{figure}

In the Mira binary system, Mira A is an evolved star that is 
undergoing a period of enhanced mass loss as it moves along the asymptotic
giant branch (AGB) on route to becoming a white dwarf. The companion star, 
previously classified as a white dwarf, now has a less clear classification
\citep{karovska05,ireland07} but is less luminous and any stellar outflow is 
comparably insignificant to Mira A in terms of mass-flux and energetics.
Observations of the Mira system (Martin et al. 2007; hereafter 
referred to as M07) revealed a comet-like tail 
in the Far Ultra-Violet (FUV) extending 2 degrees away to the North and an
arc-like structure in the South, reproduced in Figure \ref{observation}.
Based on recent dust observations 
of the AGB star R Hya \citep{ueta06} and associated simulations 
of that star \citep{wareing06b}, M07 postulated that the 
features around Mira are caused by motion through the interstellar medium 
(ISM) producing a bow shock and ram-pressure-stripped tail. Their 
postulation is consistent with the direction of Mira's proper motion 
\citep{turon93} of 225.8 milli-arcseonds per year in the direction 
187.1$^\circ$ East of North (corrected for solar motion). Further, they 
noted that at the revised Hipparcos-based distance \citep{knapp03} 
of 107 pc, the large space velocity of 130 \kms, calculated from the 
proper motion and the radial velocity \citep{evans67} of 63 \kms, 
is further consistent with the bow shock structure. M07  
concluded that the FUV emission is excited collisionally by 
the interaction of molecular hydrogen in the cool, wind wake with hot 
electrons in the post-shock gas resulting from the bow shock that also 
entrains and decelerates the wind. Based on the time taken to travel 
the observed length of the tail and assuming instantaneous
deceleration of bow shock material, they predicted the tail is 30,000
years old, consistent with maintaining the temperature of material in the
tail for FUV emission.

\cite{wareing07c} tested this hypothesis with hydrodynamical modelling.
The authors were able to reproduce the position and width of the bow shock
and the length of the tail - 4pc at D=107pc - although they were not able to 
reproduce the narrow tail and speculated that Mira could have recently 
entered the Local Bubble, a tenuous high temperature low-density region 
\citep{lallement03}. Mira would have had a small bow shock whilst travelling
through the warm neutral medium, which then expanded to its current size 
after the star entered the lower density Local Bubble. From the simulations, 
the time taken to generate a 4\,pc tail was found to be $2-3 \times 10^5$ 
years - 8x to 10x longer than M07's estimate. Unlike M07, Wareing et al.
predicted a much slower deceleration of material along the tail, meaning
a far longer time was required to generate a tail of the observed length. 
Further, since the simulated temperatures dropped rapidly along the tail, 
the driving mechanism for the FUV emission was no longer present.
Radio observations \citep{matthews08} measured the velocity of the tail 
material and confirmed Wareing et al's theory, predicting an age of the tail 
around 120,000 years in this scenario. \cite{esquivel10} explored the Local 
Bubble idea, firstly in an analytical solution and then in two-dimensional 
numerical simulations including adaptive mesh refinement. With this model, 
they were able to reproduce the narrow tail structure, whilst also reproducing 
the wide bow shock around the cometary head. However, their tail structure
is very turbulent, unlike the collimated appearance of the tail. This turbulence
is postulated to reheat the tail material and drive the FUV emission.

To date, no model has been able to reproduce the kink along the tail, nor the 
narrow collimation. Further, there is no clear mechanism to drive the FUV 
emission without turbulence dominating the structure, which is not observed. In 
this paper, the possibility of a shear flow in the ISM at an angle to Mira's 
proper motion vector is explored. In the next section, revisions to the model
are presented as a consequence of new distance estimates to Mira.  In 
Section $3$, the numerical approach is detailed along with the results in 
Section 4. We discuss the implications of our findings with respect to Mira, 
its environment and the emission mechanism in Section 5, before concluding 
the paper in Section 6.

\section{Model}

The same two-wind model as previously \citep{wareing07c} has been used,
consisting of a slow, dense AGB wind ejected from the position of the 
mass-losing star and a second wind representing the motion through 
surrounding medium. The model has also been used to successfully 
model the circumstellar structures around the evolved star R Hya 
\citep{wareing06b}, the planetary nebula Sh2-188 \citep{wareing06} and 
develop an evolutionary timeline for the interaction of AGB stars and 
planetary nebulae with the interstellar medium \citep{wareing07b}.

CO line observations \citep{ryde00} define an AGB wind mass-loss rate  
of $3 \times 10^{-7}$ \mdot\ and a velocity of 5 \kms. An unphysical 
temperature of $10^4$ K was used as the cooling curves extend no 
further but this does not affect the overall result \citep{wareing06b}.
In the direction of motion, the inner rim of the bow shock has 
been remeasured at 5 arc-minutes South of the binary system. 
Recent revisions of Mira's parallax have resulted in a new value of 
$10.91 \pm 1.22$ milli-arcseconds \citep{vanleeuwen07}, placing the 
binary system at a distance of 92\,pc $\pm$ 10.3\,pc. At this distance, 
Mira's space velocity is reduced from the original estimates to 
$116.9 \pm 8\%$ \kms. The angle of inclination of Mira's proper 
motion vector to the plane of the sky is now at 32.6$^\circ$ and the 
bow shock is at the stand-off distance of 0.134\,pc $\pm$ 
0.015\,pc. A ram pressure balance between the AGB wind and
the Local Bubble implies a Local Bubble density of $n_{\rm H}$ = $0.019 \pm 
25\%\,$cm$^{-3}$. This density depends directly on AGB wind 
mass-loss rate and velocity. The physical length of Mira's tail is approximately 
3.8\,pc. The Local Bubble temperature has been set at $10^6$\,K
\citep{lallement03}. 

Our previous model considered Mira moving through a smooth warm neutral 
medium with a low density inferred from the ram pressure balance above. We 
now postulate that at some point during the AGB phase, Mira passed into 
the low density Local Bubble from a typical warm netural medium with a density
25 times higher than that of the Local Bubble. Whilst passing through the ISM,
the bow shock would have been 5 times closer to the star. We calculate this 
ISM to have a temperatures of $10^4$\,K and a density of 
$n_{\rm H}$ = $0.475 \pm 25\%/23\%$ cm$^{-3}$ consistent with warm neutral
medium conditions typical of Mira's galactic disk location.

We evolve the simulation for enough physical time to form the long narrow 
tail whilst Mira is in the ISM and then linearly ramp the density and temperature
to the Local Bubble conditions over a period of 10,000 years as well as
linearly changing the angle of the oncoming ISM from 30$^\circ$ back 
to 0$^\circ$ with respect to the $x$-axis of the simulation domain. This
angle has been directly estimated from the observation in Figure \ref{observation}. A 
10,000-year timescale aids smooth running of the simulations and 
is more realistic since the boundary between regions is unlikely to be an
instantaneous change.

\section{Simulations}

We have employed an improved version of the {\sc cubempi} code as used
previously. We took a numerical domain of $500\times300\times300$ 
cells with the central star placed at cell coordinates 
($x$,$y$,$z$) = (50,150,150). Each cell is a regular cube 0.01\,pc 
on a side giving a physical domain of 5\,pc $\times$ 3\,pc $\times$ 3\,pc. 
The numerical scheme is second order accurate and based on a conservative
upwind Godunov-type shock-capturing
method developed by \cite{falle91} using a Riemann solver due to 
\cite{vanleer79}. The homogeneous stellar wind, ISM and Local Bubble are treated
as ideal gases. The method also includes the effect of radiative cooling above 
10$^4$\,K via parameterized cooling curves calculated from 
\cite{raymond76}, although ionisation state is not tracked. 
Mass-loss is effected by means of artificially resetting the 
hydrodynamical variables at the start of every timestep in a 
volume-weighted spherical region of radius 5 3/4 cells centred on
the position of the central star. The initial grid is otherwise entirely
filled with ISM conditions and ISM in-flow is effected by an in-flow
boundary condition at $x=0$. The change to Local Bubble conditions
is achieved by linear alteration of these in-flow boundary conditions.
We ran a number of simulations exploring the parameters discussed
above, in particular the extremes of the parallax error. 
We also investigated a number of angles (10$^\circ$, 20$^\circ$, 
30$^\circ$ \& 40$^\circ$) between the proper motion of Mira and 
the oncoming ISM flow. The results of this sensitivity study are
consistent and the data presented in the next section represent
the best fit to the observations.

\section{Results}

\begin{figure}
\begin{center}
\epsscale{0.30}
\plotone{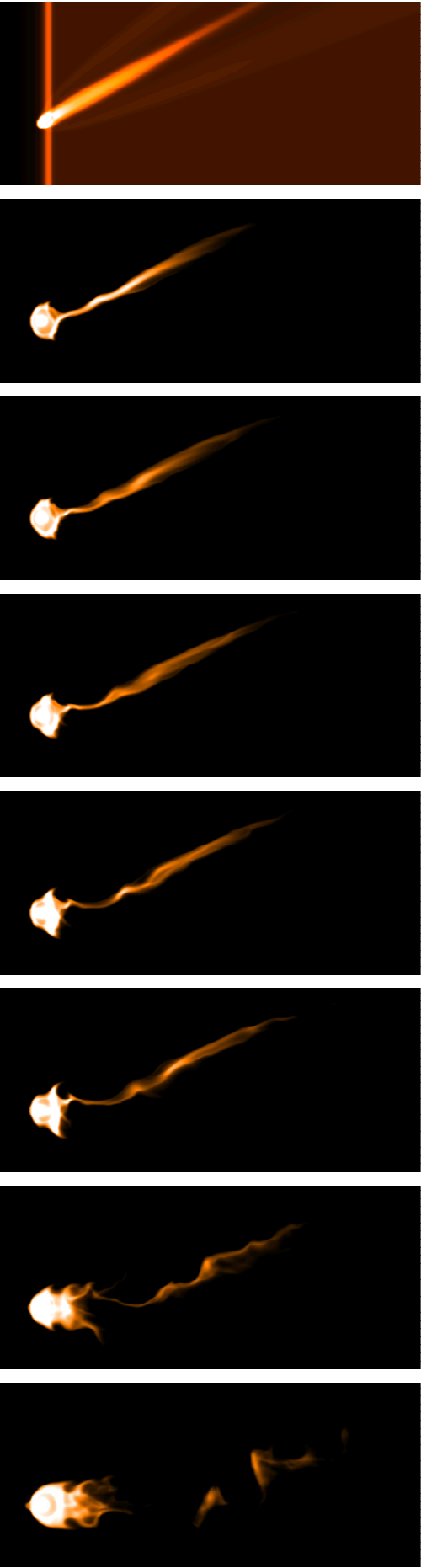}
\caption{Simulated evolution of the Mira system. The density datacube has been
collapsed perpendicular to the direction of motion. The top row shows Mira just 
entering the local bubble. Descending rows are then 75, 87.5, 100, 112.5, 125, 
175 and 225 kyrs after this time. Each panel is 2.25pc\,$\times$\,5pc.}
\label{density}
\end{center}
\end{figure}

The simulations begin at the onset of AGB mass-loss and are performed in
the frame of reference of the star.  The wind from the star forms into a 
bow shock upstream, positioned at a stand-off distance which can be 
understood in terms of a ram pressure balance between the stellar wind 
and the ISM. Simulated temperatures at the head of the bow shock are 
in agreement with strong shock theory: T $\sim (3/16)mv^2/k$. Ram-pressure-stripped 
material from the head of the bow shock forms a tail behind the nebula 
at an angle of 30$^\circ$ to the $x$-axis. After 250,000 years, the tail is 
narrow and remains collimated to the edge of the simulated domain - a 
distance down the tail of ~4\,pc. We show this instance in time in the top 
row of Figure \ref{density}, where the passage of 
the Local Bubble is clear. We then show the collapsed density datacube  
at 75,000, 87,500, 100,000, 112,500, 125,000, 175,000, 
and 225,000 years after crossing the boundary. As the Local 
Bubble conditions sweep past the location of Mira, the bow shock 
expands to the new stand-off distance five times further away from 
the star after 50,000 years.

Given that the ambient medium temperature has increased by a factor 
of 100, and the density decreased by only a factor of 25, the relative 
thermal pressure in the Local Bubble region is four times higher than 
that in the ISM. This higher pressure has begun to compress the tail, 
increasing density and temperature (panel 2, 75,000 years after Mira 
entered the Local Bubble). 100,000 years after Mira has entered the 
Local Bubble, the simulated structure compares very well to the
observed one. The bow shock has the correct position and width. The 
tail has been narrowed by the pressure of oncoming material. Whilst
the simulation does not reproduce the gap in the tail, the observed 
position and angle of the kink in the tail is best fit by the simulation at 
this time. The simulation predicts Mira entered the Local Bubble 
$100,000 \pm 25,000$ years ago. After this time, shown further down 
Figure \ref{density}, the position of the kink 
moves down the tail, going out of agreement with the observation. 
After 175,000 years the pressure driving the initial compression and 
reheating has now begun to destroy the tail. Material from the bow 
shock is now forming a wide, cool turbulent wake and no longer
resembles a collimated tail.

\section{Discussion}

\subsection{Pressure balance?}

Setting to one side the question of whether the Local Bubble exists, it is 
thought to be an ancient supernova driven bubble approximately 100\,pc 
across and 1-10 million years old. The outward driving pressure is likely 
to be very low, hence our choice of a factor 4 pressure increase. The 
Local Bubble in the simulations of \cite{esquivel10} is at a pressure of 50 times
greater than their warm neutral medium, high for an ancient and 
broad structure. This unrealistically large pressure difference is likely to 
drive the strong ram-pressure-stripping and turbulent tail conditions seen 
in their model. It is of interest to consider what would happen if the warm 
neutral medium and the Local Bubble had reached thermal pressure 
equilibrium.  To maintain the temperature ratio of 100, the density ratio 
for thermal equilibrium must be 0.01. With the density in the Local 
Bubble fixed by the position of the bow shock, the warm neutral medium 
density must then be $n_{\rm H}$ = 1.9 cm$^{-3}$. The results of a 
simulation with these conditions shows similar, but more rapid, evolution to the 
one presented in Figure \ref{density}. The tail material is swept more rapidly 
behind the star and after 75-100,000 years the kink has all but disappeared. 
Temperature also reduces drastically down the length of the tail and more 
quickly as the material is entrained behind Mira. After 150,000 years, it is 
ten times cooler than the bow shock. These results
would suggest that the Local Bubble is not yet in pressure equilibrium with 
its surroundings.

\subsection{The origins of the flow in the ISM}

This work suggests a flow in the ISM at $30^\circ$ to Mira's proper motion
vector is responsible for the kink in the tail. The passage into the Local Bubble
leads to compression and reheating of the tail material via 
an over-pressure compared to the surrounding warm neutral medium.
Practically no implications of a flow in the ISM exist in the literature; 
\cite{ueta10} imply a bow shock around the AGB star R Cas
has been compressed on one side by a flow in the ISM, but such one-sided
emission can be more easily understood in terms of instabilities seen in AGB
star bow shocks, evidence for which has been detected in several objects,
specifically Sh\,2-188 \citep{wareing06}, in Mira itself, launched from the bow shock
leading to the observed ring in the tail and also clearly in the instability-dominated bow
shocks recently revealed around X Her and TX Psc \citep{jorissen11}. 
Whilst it is possible that Mira's tail is entirely instability driven, in the context of
our previous simulations of AGB star tails and associated instabilities and vortices 
\citep{wareing07, wareing07b} it has been impossible to reproduce
the distinct directional change of the tail.
The $30^\circ$ angle implies the flow speed is around 1/3 of Mira's own velocity,
or approximately 40 \kms. The sound speed of the warm neutral medium
at 10$^4$\,K is approximately 15\kms, implying this is a Mach 2 to 3 flow.
A possible origin then, supported by the indications above that the Local Bubble
is probably still pressure driven, is that the shell of ISM material swept-up
by the expanding supernova remnant is causing the flow.

\subsection{Emission}

One central issue is the source of the long-lived FUV tail emission. M07 modelled
several options with CLOUDY, and following a comparison to GALEX grism mode
observations, only H$_2$ emission and (only marginally) isothermal 10$^5$ K plasma 
emission were consistent with the lack of NUV and H$\alpha$ emission. M07 thus 
concluded that the FUV emission is excited collisionally by the interaction of H$_2$ 
in the cool wind wake with hot electrons in the post-shock gas as discussed previously.
However, since then, both simulations and further observations have found the tail to be 
4 to 10 times older. This raises the difficulty of maintaining sufficient thermal energy over 
the extended lifetime and  length of the tail in order to power the FUV emission. Esquivel 
et al implied turbulent shock heating may lead to the UV emission, but the 
high level of turbulence is likely to be caused by the unrealistically high pressure
difference between neutral medium and Local Bubble conditions. The power source of
the FUV emisison has remained unclear.

\begin{figure}
\begin{center}
\epsscale{0.30}
\plotone{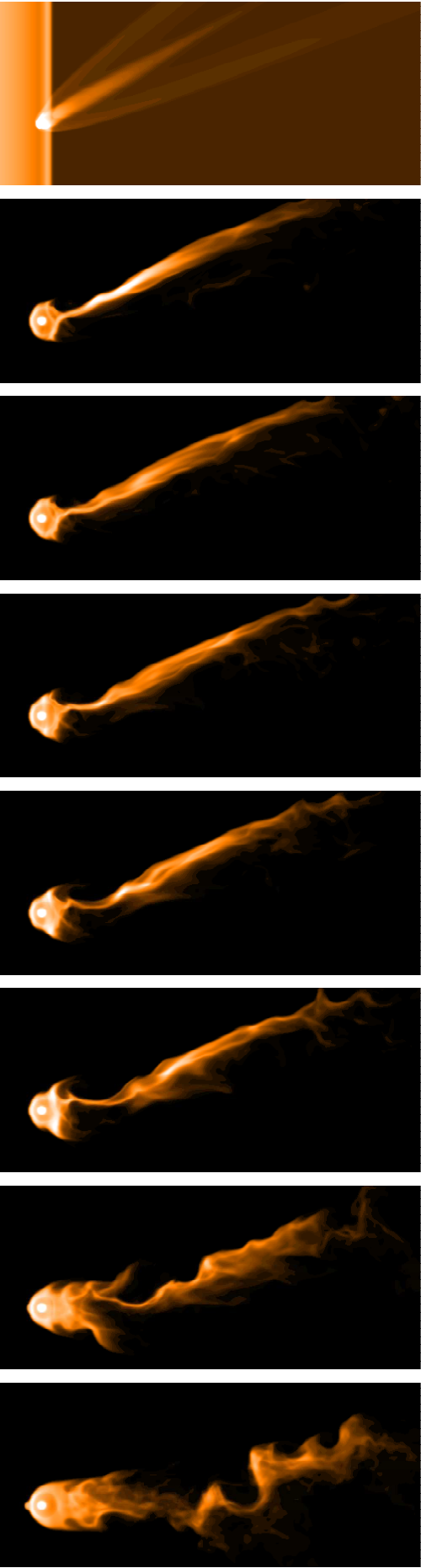}
\caption{Simulated logarithmically-scaled emission from the Mira system. The top
row shows Mira just entering the local bubble. Descending rows are then at the 
same times as Figure \ref{density} Each panel is 2.25pc\,$\times$\,5pc.}
\label{emission}
\end{center}
\end{figure}

In previous simulations without a shear, we have found that the temperature drops along 
the tail far too quickly to support the existence of hot electrons - see the first row
of Figure \ref{emission}. However, in the shear model, with the hot 
local bubble, the kinked tail is exposed to hot oncoming local bubble material, compressing
and reheating the tail material. We calculate naive emission from the tail based on free-bound hydrogen
emission, $E \propto n_e n_p k T_e$, assuming $n_e = n_p = \rho$ and $T_e = T$.
Figure \ref{emission} thus shows $\rho^2\,T$ at equivalent times to the 
density plots. We suggest that the FUV emission is thermally driven by the compression effect of 
the local bubble material, rather than bow-shock heating. Consequently,
when the tail is eventually swept behind the star - 250,000 years after entering the local 
bubble in our simulations - the recompression effect ceases and the FUV emission will switch off. 
It should be noted though that this proposed reconciliation of the source 
of emission on a timescale that agrees with the age estimates of Matthews et al. is achieved 
primarily through choice of Local Bubble conditions and shear flow angle and can be adjusted
as such. Radiative transfer modelling, beyond the scope of this work, is required to examine 
this in any further detail.

\subsection{Still one of a kind}

Observationally, the tail of Mira still remains unique. However, all 
evolved AGB stars show similar {\it or stronger} winds to Mira
and several bow shocks have now been observed around similar
stars. The question still remains then as to whether Mira is a 
proto-type or an exception, regarding its tail. 
There are now two main differences between Mira 
and typical similar stars. Specifically, the high space velocity and the
transit from a warm neutral ISM with a shear flow into the Local
Bubble. These conditions combine to make Mira's tail observable
only briefly and also imply that tails in more common ISM conditions
are unlikely to generate sufficient recompression and heating to 
produce FUV emission from their tails.

Tails behind AGB stars may therefore be common and in most 
cases considerably wider. However, the particular set of conditions
that we find have combined to make Mira's tail collimated over a great length and observable, 
are unlikely to be commonly observed in the FUV.  Shorter, wider, turbulent wakes are much more
likely to be observed.

\section{Conclusion}

We have shown in this paper that the kinked nature of Mira's tail
can be reproduced by modelling the star's passage through a warm
neutral medium with a flow at an angle of 30$^\circ$ to the proper 
motion vector. Mira then enters a hot low density environment, 
thought to be the Local Bubble, where the bow shock expands to its 
current position and the tail is compressed and reheated, driving
collisionally excited FUV emission in agreement with the GALEX observations
\citep{martin07}. These simulations predict Mira entered the Local Bubble 
$100,000 \pm 25,000$ years ago, in agreement with the radio observation 
estimate of \cite{matthews08}. The discrepancy between the source of
energy driving the FUV emission and the age of the tail has therefore been
resolved.

Mira's tail is a transient structure - only visible for the 100,000 years or so
between the beginning of the compression and reheating following entry
into the Local Bubble and the destruction of the tail by the formation of
a turbulent wake. Without the high temperatures of the Local Bubble, FUV
emisison would be unlikely making us lucky indeed to be able to observe 
wonderful Mira at this spectacular and unique time in its evolution.

\acknowledgments

The author acknowledges useful discussions with Albert Zijlstra and Sam 
Falle and also constructive remarks from an anonymous referee that led 
to many improvements to the original manuscript. The numerical 
computations were carried out using the STFC-supported UKMHD 
computing facility at the University of Leeds.

\clearpage

\clearpage

\end{document}